# NON EXTENSIVITY IN METEOR SHOWERS

Oscar Sotolongo-Costa[1], R. Gamez[2], F. Luzón[1,3,4], A. Posadas[1,3,4], Pablo Weigandt Beckmann[5]
1.-Cátedra "Henri Poincaré" de Sistemas Complejos, Universidad de La Habana.
2.- Departamento de Física Aplicada, Universidad de Almería.
3.- Instituto Andaluz de Geofísica. Granada. Spain.
4.-Instituto de Geofísica y Astronomía, ACC; Cuba
5.- Departamento de Física, Universidad Autónoma de Chiriquí, Panamá

**Abstract**:
A meteor shower is a luminous phenomenon that takes place by the entry into the Earth's atmosphere of a cascade of particles coming from a stream intersected by our planet in its orbit.
Here we investigate the possibility of a description of the mass distribution of meteoroids in meteor showers in terms of a non extensive formulation, which could shed light and give some insight into the origin of such particles.

## *I. – Introduction*

Meteor Showers (MS) are flurries of meteors seemingly emanating from spots of the sky at particular times of the year. This happens when the Earth intersects some of the particle swarms that move around the Sun.

As to the origin of MS, the hypothesis of its cometary origin is widely accepted.

Comets are celestial bodies formed by frozen dust and gas. When these objects approach their perihelion, the interaction of solar wind with its surface causes the sublimation of its components. As a result, comets produce a long tail always in opposite direction to the Sun. The tail is made up by a large amount of particles which eventually spread out along the entire orbit of the comet forming a meteoroid stream. When our planet intersects the orbit of the comet, meteoroids fall into Earth's atmosphere at high speed ionizing it and causing a shower of luminous particles

According to [1], the tail of the comet is subjected to the periodic influence of Jupiter, so that the particles of the tail form a 'meteoric pipe' that stretches continuously. The gravity forces act on the comet and its swarm, so that each expulsion of particles becomes more and more unique and separates from the comet; this originates a complex system of meteoric fibres inside the pipe.

The spatial distribution of such fibres and the way the Earth intersects one or more of them determines the duration and intensity of showers. The meteoroids entering the Earth's atmosphere produce different luminous flares, according to their mass and the relative speed of the swarm.

Meteors seem to radiate from a single point in the sky. The radiating point (or just "the radiant") is caused by the effect of the perspective on the observer located on Earth's surface. The showers are usually named after the constellation in which their radiant lies at the time of maximum stream. Such is the origin of names like Perseids, Leonids, etc.
Apart from the questions raised concerning the origin of MS, features such as their size distribution are of interest to get some clue about the processes which could have originated these particles.

It is worth mentioning that the study of the luminous flux of meteoric particles indicates that they are distributed following a power law [2, 3]. The above fact, the existence of complex interactions among the meteoroids, the interaction of the comet with solar wind, its fragmentation and many other phenomena present on this celestial effect constitute a very complex field of study and speculation, posing, among other things, the question of how the observation of luminous intensities in MS can give us information about the characteristics of MS and their generators, the comets, since MS are a manifestation of the interaction between Earth's atmosphere and the cometary particles. As far as we know, no attempt has been made to deduce a functional dependence related to the mass distribution function of meteoroids, starting from first principles.

As the luminous distribution of meteoroids can be deduced from observations, and on this basis their size distribution can be inferred, more information could be obtained by observations if the size distribution function (SDF) could be linked to a given theoretical framework. Taking the above into account, we will expound, on the basis of physical grounds, a description of SDF caused by fragmentation phenomena. If the observed SDF can be fitted with theoretical expressions obtained from specific starting hypotheses, maybe we could say something about its possible origin.

On the other hand, if there are strong suspicions that the meteoroids are the result of violent fragmentation processes, we can validate our theoretical framework.

## *II. - Non-extensivity in fragmentation*

As a result of developments in materials science, combustion technology, geology and many other fields of research, there has been an increase of interest in the question of object fragmentation.

Some attempts have been made to derive the fragment size distribution function from the maximum entropy principle [5, 6], subjected to some constraints which mainly came from physical considerations about the fragmentation phenomena. The resulting fragment distribution function describes the distribution of sizes of the fragments in a regime in which scaling is not present.

In the general field of fragmentation there is a collection of papers [4-7] where a transition occurs from a "classical" distribution of fragments (*e.g.* log-normal or Rossin-Ramler-like) to a power law distribution. This transition has not been adequately explained in terms of any general principles, although in [8] the representation of the fragmentation process, in terms of percolation on a Bethe lattice. leads to a transition to a power law in the distribution of fragment sizes. Other theoretical efforts like the study of dimensional crossover in fragmentation of thick clay plates and glass rods [10], and very interesting experiments that show scaling in violent fragmentation processes, like the burst of a fuel droplet [9] are also proofs of the interest in these phenomena.

Because scaling certainly occurs when the energy of the fragmentation process is high, this suggests that the traditional statistical analysis is only applicable to low energies. However, we believe that the maximum entropy principle is completely universal and has an almost unlimited range of applications. Consequently, we would expect to be able to use in describing the transition to scaling as the energy of the fracture grows.
The expression for the Boltzmann-Gibbs entropy $S$ (*e.g.* Shannon's form) is given by

$$S = -k \sum_{i=1}^{W} p_i \ln p_i \qquad (1)$$

where $p_i$ is the probability of finding the system in the microscopic state $i$, $k$ is Boltzmann's constant, and $W$ is the total number of microstates. This has been shown to be restricted to the domain of validity of Boltzmann-Gibbs (BG) statistics.

These statistics seem to describe nature when the effective microscopic interactions and the microscopic memory are short ranged [8]. The process of violent fractioning, like that of droplet micro-explosions in combustion chamber, blasting and shock fragmentation with high energies and many others, leads to long-range correlations between all parts of the object being fragmented.

Fractioning is a paradigm of non extensivity, since the fractioning object can be considered a collection of parts which, after division, have an entropy larger than that of their union *i.e.*, if we denote by $A_i$ the parts or fragments in which the object has been divided, its entropy $S$ obeys $S(\cup A_i) < \Sigma_i S(A_i)$, defining a "superextensivity" in this system. This suggests that it may be necessary to use non-extensive statistics, instead of the BG one. This kind of theory has already been proposed by Tsallis [8], who postulated a generalized form of entropy, given by

$$S_q = k \frac{1 - \int_0^\infty p^q(x) dx}{q-1} \tag{2}$$

The integral runs over all admissible values of the magnitude $x$ and $p(x)dx$ is the probability of the system being in a state between $x$ and $x+dx$. This entropy can also be expressed as

$$S_q = \int p^q(x) l_q p(x) dx \tag{3}$$

where the generalized logarithm $l_q(p)$ is defined in [8] as

$$l_q(p) = \frac{p^{1-q} - 1}{1-q} \tag{4}$$

where $q$ is a real number. It is straightforward to see that $S_q \to S$ when $q \to 1$, recovering BG statistics.

Here, we try to deduce the size distribution function of meteoroids starting from first principles of physics, *i.e.*, maximum entropy principle with Tsallis's entropy formalism. Fracture processes characterized by a strong correlation among all parts of the body must be described from the non extensive statistics [8].

Let us start from the Tsallis's entropy for the mass distribution of the fragments in the form

$$S_q = k \frac{1 - \int_0^\infty p^q(M) dM}{q-1} \tag{5}$$

where $M$ is a non-dimensional mass.

Let us impose the normalization condition

$$\int_0^\infty p(M) dM = 1 \tag{6}$$

and a "$q$-conservation" of mass in the form

$$\int_0^\infty M p^q (M) \, dM = 1 \tag{7}$$

From the maximum entropy principle, SDF can be determined in a straightforward manner. The problem is to find the maximum entropy conditioned by the normalization of the probability distribution and the $q$-conservation of mass already declared. Using the method of Lagrange multipliers the following functional can be constructed:

$$L(p;\alpha;\beta) = \frac{S_q}{k} - \alpha \int_0^\infty p(M) \, dM + \beta \int_0^\infty M \, p^q (M) \, dM \tag{8}$$

Its extremization leads to

$$p(M) \, dM = a (1 + bM)^{-\frac{1}{q-1}} dM \tag{9}$$

where $a$ and $b$ are $q$ dependent constants that can be used as adjustment parameters. This is the method we will follow to find an expression for the SDF of meteoroids and contrast it with observations of MS.

It is worth to say that (9) gives, in the asymptotic limit, $p(M) \sim M^{-n}$ with $n = 1/(q-1)$. *i.e*, scaling is obtained in this formulation as one of its essential characteristics.

## *III.- Method*

Meteor showers are subjected to visual observation, where the visual magnitude of meteoroids is registered. From this, the luminous flux and the mass of the particles in the shower can be determined. In this case we are specially interested in the mass and its relation to the visual magnitude.

The mass of the meteoroids can be obtained by [12]:

$$M \sim e^{-\gamma m} \tag{10}$$

where $m$ is the visual magnitude of the particle.

For a description of the resulting SDF produced in the comets for its interaction with the solar wind we analyse SDF of several MS, *i.e.,* Leonids, Perseids and Lyrids (Table I). It is known that these MS are produced by swarms emerging from a comet.

*Table I MS processed and their original comets.*

| Meteor Shower | Original Comet | Years | # observations |
|---|---|---|---|
| Leonids (LEO) | Tempel Tuttle | 1995 - 1999 y 2001 | 8990 |
| Perseids (PER) | Swift Tuttle | 1995 - 1999 y 2001 | 8095 |
| Lyrids (LYR) | Thatcher 1861 I | 1995 - 1999 y 2001 | 786 |

Data were taken from [13]. As atmosphere restricts visualizations of low intensity meteoroids, to determine the number of particles entering the atmosphere a correction factor was introduced to account for the perception probability of each magnitude [14].

Here we start from the idea that mass distribution in MS may exhibit similar characteristics as fragment distribution functions emerging from fragmentation processes with high energy.

As mentioned above, fragmentation processes have been recently formulated on the ground of Tsallis' non extensive statistics. Fragmentation can occur as a violent process, where long range interactions are present in the fragmenting body, so that the extensive statistics of Boltzmann-Gibbs cannot be applied. Tsallis' formalism gives, as we already have seen, the mass distribution function (8). Now we use the relation (10) to get

$$\frac{dM}{dm} = ce^{-\gamma m} \quad (11)$$

where $c$ and $\gamma$ are constants. Taking into account that

$$p(m)dm = p(M)dM \quad (12)$$

we finally get

$$p(m)dm = \mu e^{-\gamma m} \left( k + \rho e^{-\gamma m} \right)^{-\frac{1}{q-1}} dm \quad (13)$$

This is the expression we´re going to use to compare with the data already mentioned. It seems worthless to be concerned about the meaning of the specific values of the fitting constants, since what really matters in this case is the physical considerations leading to the above followed method. Namely, the fact that the fragments forming the comet and detached by solar wind come from violent processes of fragmentation. As a result, their mass distribution obeys (8) and visual magnitude distribution corresponds to (13).

## IV. - Application of the model to MS

In figures 1 to 3 the dots show the data of the visual magnitude of the MS Leonids, Perseids and Lyrids. The fitting curves correspond to equation (13). In all cases, the agreement with observational data is pretty good.

The results of the fitting highlights the general character of non-extensivity for the phenomenon of MS. We can assume that meteoric particles exhibit the same characteristics as when they were in the interior of their carrier comet, *i.e,* assume that the observed mass distribution is the same as that of the solid material forming the comet's nucleus. According to this hypothesis the genesis of such particles occurred before the 'birth' of the comet. They are liberated from the comet's nucleus when it approaches the Sun, forming a complex swarm of particles along the comet's orbit.

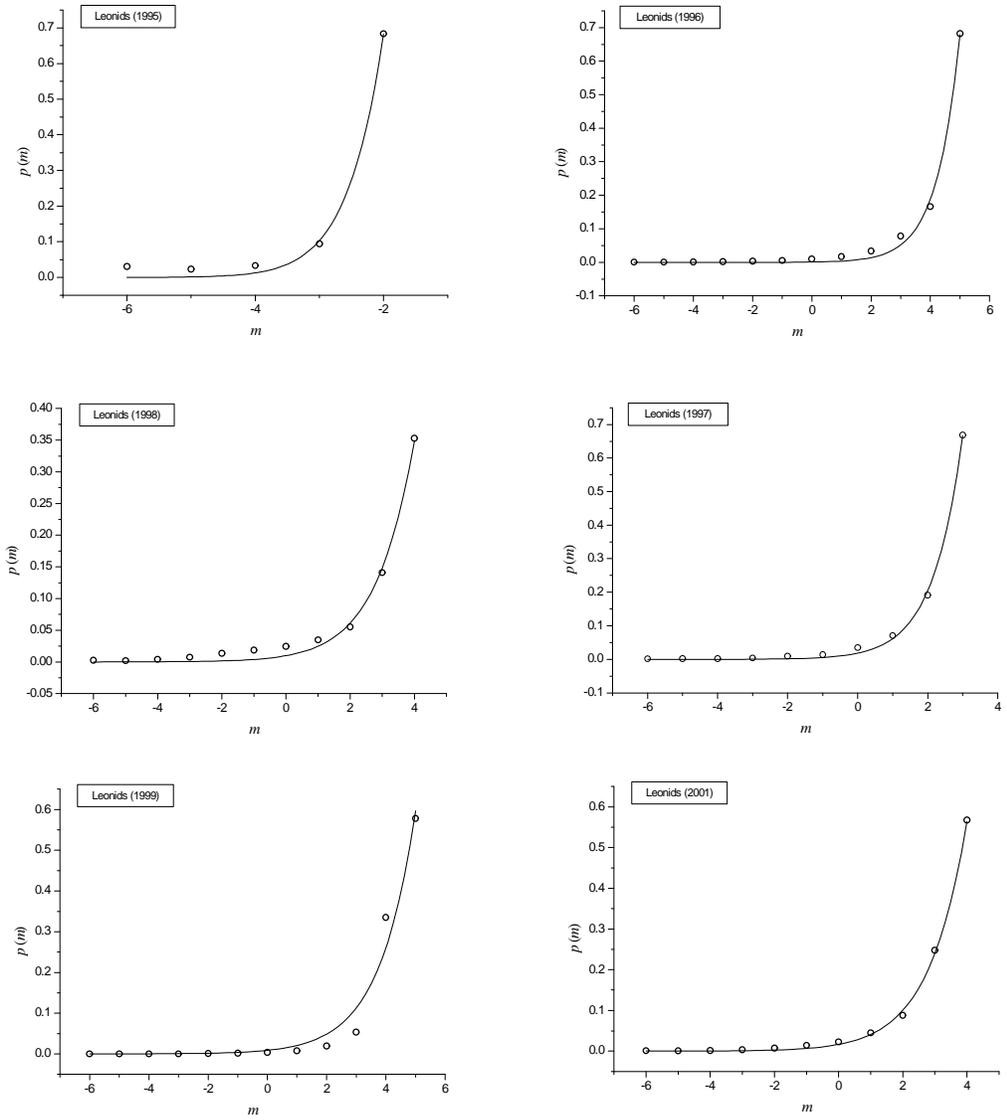

*Fig.1 – Density distribution of meteoroids by magnitude in Leonids for 1995-1999 and 2001; fitting was made with equation (13).*

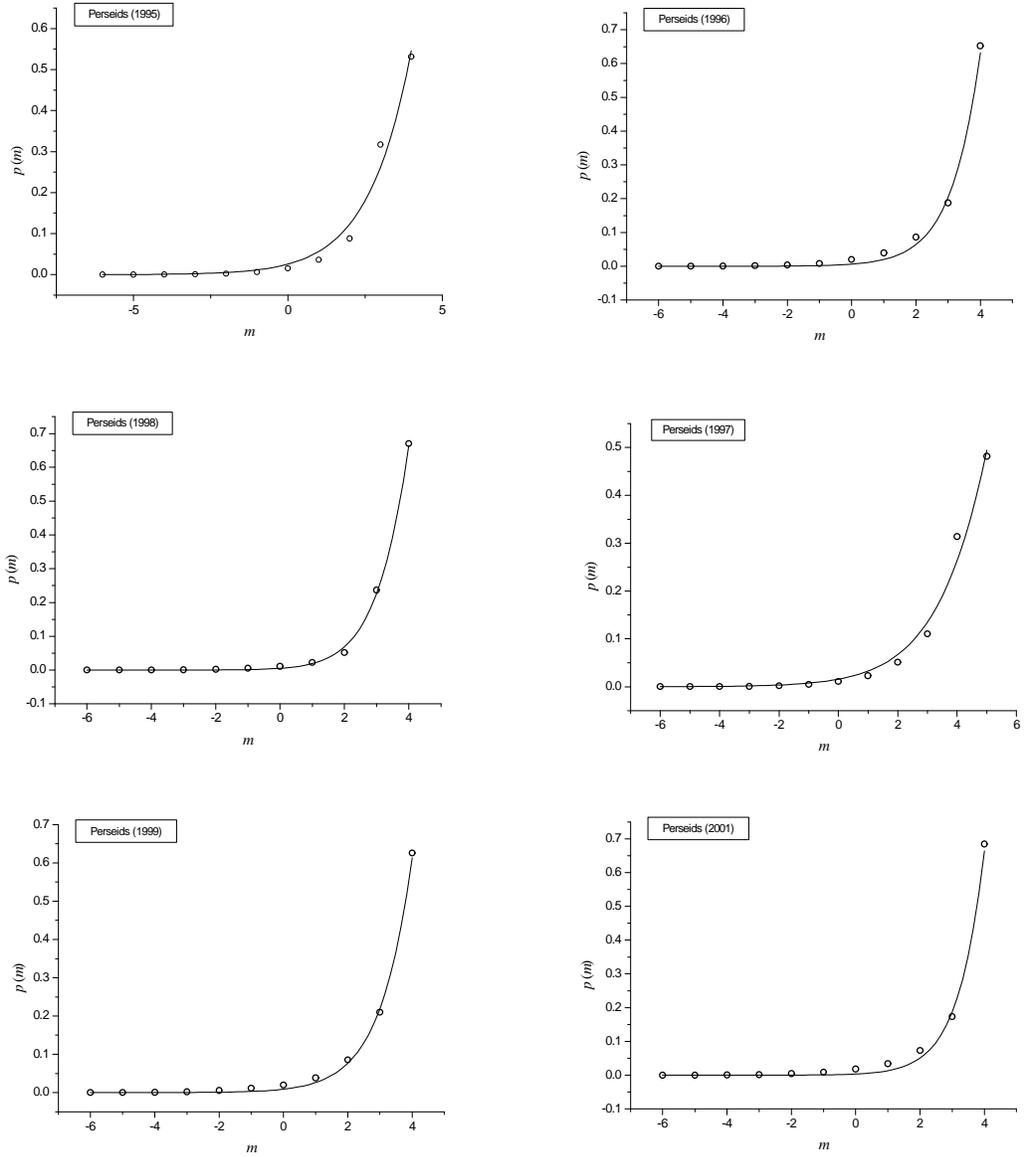

*Fig.2 – Density distribution of meteoroids by magnitude in Perseids for 1995-1999 and 2001; fitting was made with equation (13).*

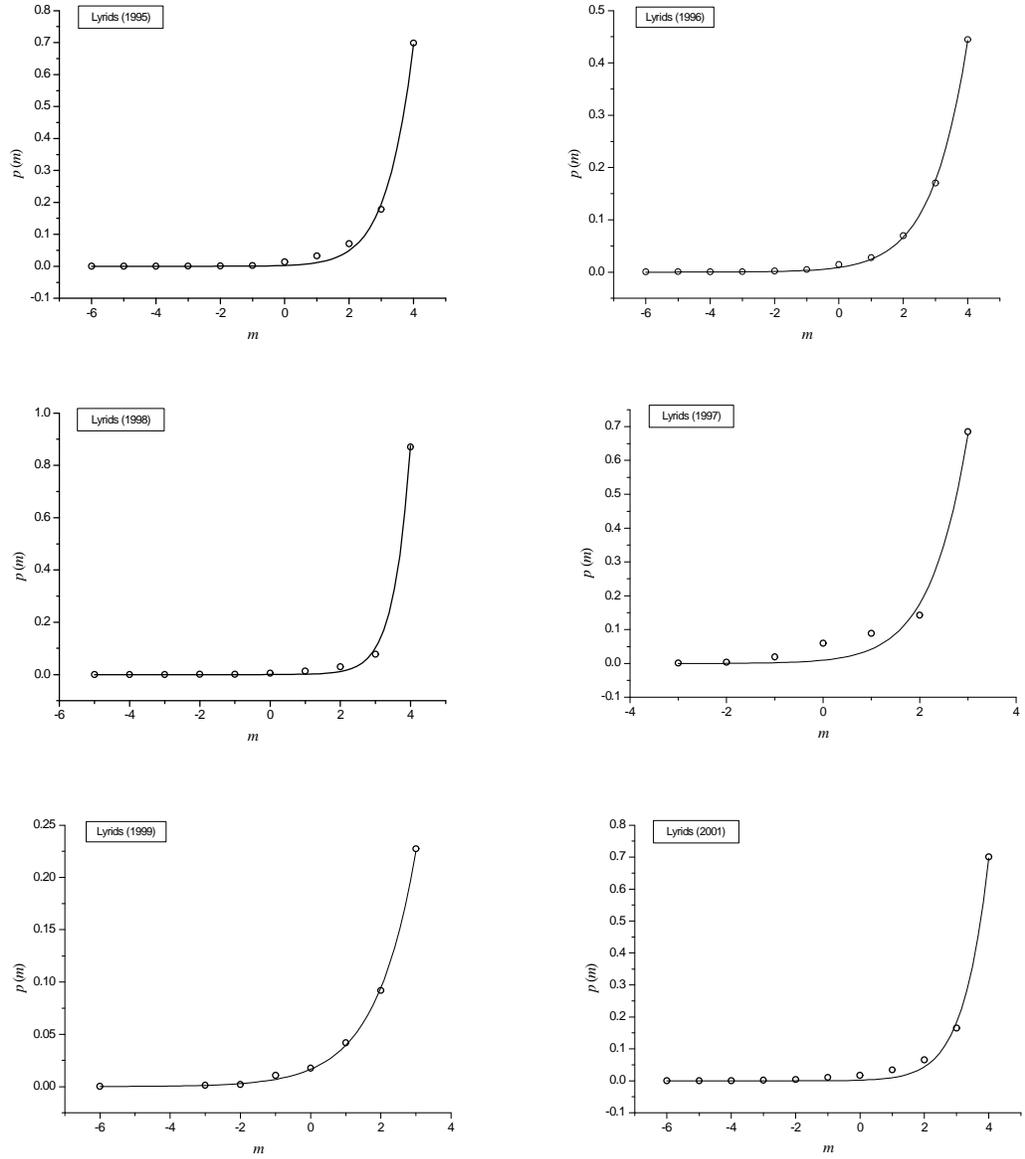

Fig.3 – Density distribution of meteoroids by magnitude in Lyrids for 1995-1999 and 2001; fitting was made with equation (13).

The particles forming the comet are the residuals of the original processes of the Solar System. In an early phase, solid bodies of different sizes were formed. It is widely accepted that during such processes the interactions among these bodies led to frequent and violent collisions. We assume that such early processes of fragmentation gave origin to the size distribution we presently observe. The fitting with (13) leads to think that such processes were characterized by long range correlations, *i.e*, violent breaking phenomena.

## V. - Conclusions

First of all, we see once again that non-extensive statistics using Tsallis' entropy can be applied, in a satisfactory fashion, to phenomena related to violent fragmentation processes. If we use Shannon – Boltzmann's entropy

$$S = -\int_0^\infty p(M)\ln(p(M))dM$$

subject to the conditions

$$\int_0^\infty p(M)dM = 1 \quad \text{and} \quad \int_0^\infty Mp(M)dM = \langle\langle M \rangle\rangle$$

we get, by means of Lagrange variational method,

$$p(M)dM = Ce^{-\lambda M}dM$$

and, because of (13),

$$p(m)dm \sim e^{-\left(\eta e^{-\gamma m} + \gamma m\right)}dm.$$

This relation, in general, does not fit with the whole data, showing that Boltzmann's Statistics fails in this case.

It is safe to conclude that in MS the universal character of fragmentation processes is confirmed, as well as the generality of the non extensive entropy formalism.

Nevertheless, old swarms, due to the loss of their particles along the orbit for many years, can lead to deviations from the power law distribution. This is due to the Poynting effect: the drift of the smaller particles by solar wind. Older swarms are, therefore, more affected than newer ones. The antiquity of the swarms intersecting terrestrial orbit should have influence on the mass distribution function of MS.

Consequently, future works could be addressed to the question of the relationship between the age of the swarm and the mass distribution of the corresponding MS.

## Acknowledgements

We wish to acknowledge the collaboration of Dr. E. Weigandt for helpful suggestions. One of us (O.S.) also wishes to acknowledge the warm hospitality of all the people related to UNACHI during his time as visiting professor.